\documentclass[prl,fleqn,twocolumn,amsmath,amssymb]{revtex4}
\pdfoutput=1
\usepackage{graphicx}
\usepackage{color}
\usepackage{graphicx}
\usepackage{dcolumn}
\usepackage{bm}
\usepackage{amsmath}
\usepackage{amsfonts}
\usepackage{bbm}
\usepackage{subfigure}
\usepackage{setspace}
\usepackage{pxfonts}
\usepackage{pstricks}

\newcommand{\beq}{\begin{eqnarray}}
\newcommand{\eeq}{\end{eqnarray}}

\newcommand{\bmp}{\noindent\begin{minipage}{16cm}}
\newcommand{\emp}{\end{minipage}\vskip 7mm} 

\def\drawbox#1#2{\hrule height#2pt
        \hbox{\vrule width#2pt height#1pt \kern#1pt
              \vrule width#2pt}
              \hrule height#2pt}

\def\Asym#1#2{\vcenter{\vbox{\drawbox{#1}{#2}
              \kern-#2pt 
              \drawbox{#1}{#2}}}}

\begin{document}
\title{\Large  \color{red}  The Conformal Window and Walking Technicolor}
\author{ $^{\color{blue}{\varheartsuit}}$Thomas A. {\sc Ryttov}}
\email{ryttov@nbi.dk}
\affiliation{
$^{\color{blue}{\varheartsuit}}$Niels Bohr Institute, Blegdamsvej 17, DK-2100 Copenhagen \O, Denmark }



\begin{abstract}
I discuss recent progress in the uncovering of the phase diagram of non-supersymmetric gauge theories. The nature of the conformal window for higher dimensional representations suggests a possible way to construct realistic technicolor models. I then explicitly provide two such theories. One of these models also has a natural cold Dark Matter candidate.
\end{abstract}

\maketitle

The origin of electroweak symmetry breaking is one of the most outstanding problems of today in high energy physics. In the past many different ideas have been proposed to explain the mass generation of the electroweak gauge bosons with technicolor \cite{Weinberg:1979bn} being one of the best motivated extensions beyond the Standard Model (SM). Despite the elegance of the technicolor proposal it is only very recent that viable specific models not at odds with experiments have been constructed (for a review see \cite{Sannino:2008ha}).

Typically one is faced with the problem of constructing a technicolor theory that does not give a too large contribution to the $S$ parameter \cite{Peskin:1990zt} while at the same time exhibits walking dynamics \cite{Holdom:1981rm}. In the original technicolor proposal the fermions were taken to be in the fundamental representation and hence one was at odds with the Electroweak Precision Tests since a large number of fermions was needed in order to obtain the desired dynamics.

However, with recent advances in the understanding of the phase diagram of gauge theories involving fermions in arbitrary representations of the gauge group \cite{Sannino:2004qp,Dietrich:2006cm,Ryttov:2007sr,Ryttov:2007cx} new directions and possibilities for model building have been opened and envisioned \cite{Sannino:2004qp,Ryttov:2008xe}. Already a large amount of work has been done ranging from the study of Beyond SM phenomenology \cite{Foadi:2007ue}, Unification \cite{Gudnason:2006mk} and the finite temperature phase transition \cite{Cline:2008hr} together with Cosmology \cite{Nardi:2008ix}. Also the lattice is starting to probe the (near) conformal dynamics of the simplest models \cite{Catterall:2007yx}.

\section{The Phase Diagram} Let us first set the notation by denoting the generators of the gauge group in the
representation $r$ by $T_r^a,\, a=1\ldots N^2-1$. They are normalized according to
$\text{Tr}\left[T_r^aT_r^b \right] = T(r) \delta^{ab}$ while the
quadratic Casimir $C_2(r)$ is given by $T_r^aT_r^a = C_2(r)I$. The
trace normalization factor $T(r)$ and the quadratic Casimir are
connected via $C_2(r) d(r) = T(r) d(G)$ where $d(r)$ is the
dimension of the representation $r$. The adjoint
representation is denoted by $G$.

Let us first consider an $SU(N)$ gauge theory with $N_f(r_i)$ Dirac fermions in the representation $r_i, \ i=1,\ldots,k$ of the gauge group. To estimate the conformal window we shall employ the recently conjectured all-orders beta function for non-supersymmetric theories \cite{Ryttov:2007cx}

\begin{eqnarray}
\beta(g) &=&- \frac{g^3}{(4\pi)^2} \frac{\beta_0 - \frac{2}{3}\, \sum_{i=1}^k T(r_i)\,N_{f}(r_i) \,\gamma_i}{1- \frac{g^2}{8\pi^2} C_2(G)\left( 1+ \frac{2\beta_0'}{\beta_0} \right)} \ ,
\end{eqnarray}
with
\begin{eqnarray}
\beta_0 &=&\frac{11}{3}C_2(G)- \frac{4}{3}\sum_{i=1}^k \,T(r_i)N_f(r_i)  \ , \\
\beta_0' &=& C_2(G) - \sum_{i=1}^k T(r_i)N_f(r_i)  \ , \\
\gamma_i(g^2) &=& \frac{3}{2} C_2(r_i) \frac{g^2}{4\pi^2} + O(g^4) \ .
\end{eqnarray}
Here $g$ is the gauge coupling, $\beta_0$ is the first coefficient of the beta function and $\gamma_i(g^2)$ is the anomalous dimension of the fermion mass. One should note that for small coupling the beta function reduces correctly to the two loop beta function.

\begin{figure}[h]
\resizebox{8cm}{!}{\includegraphics{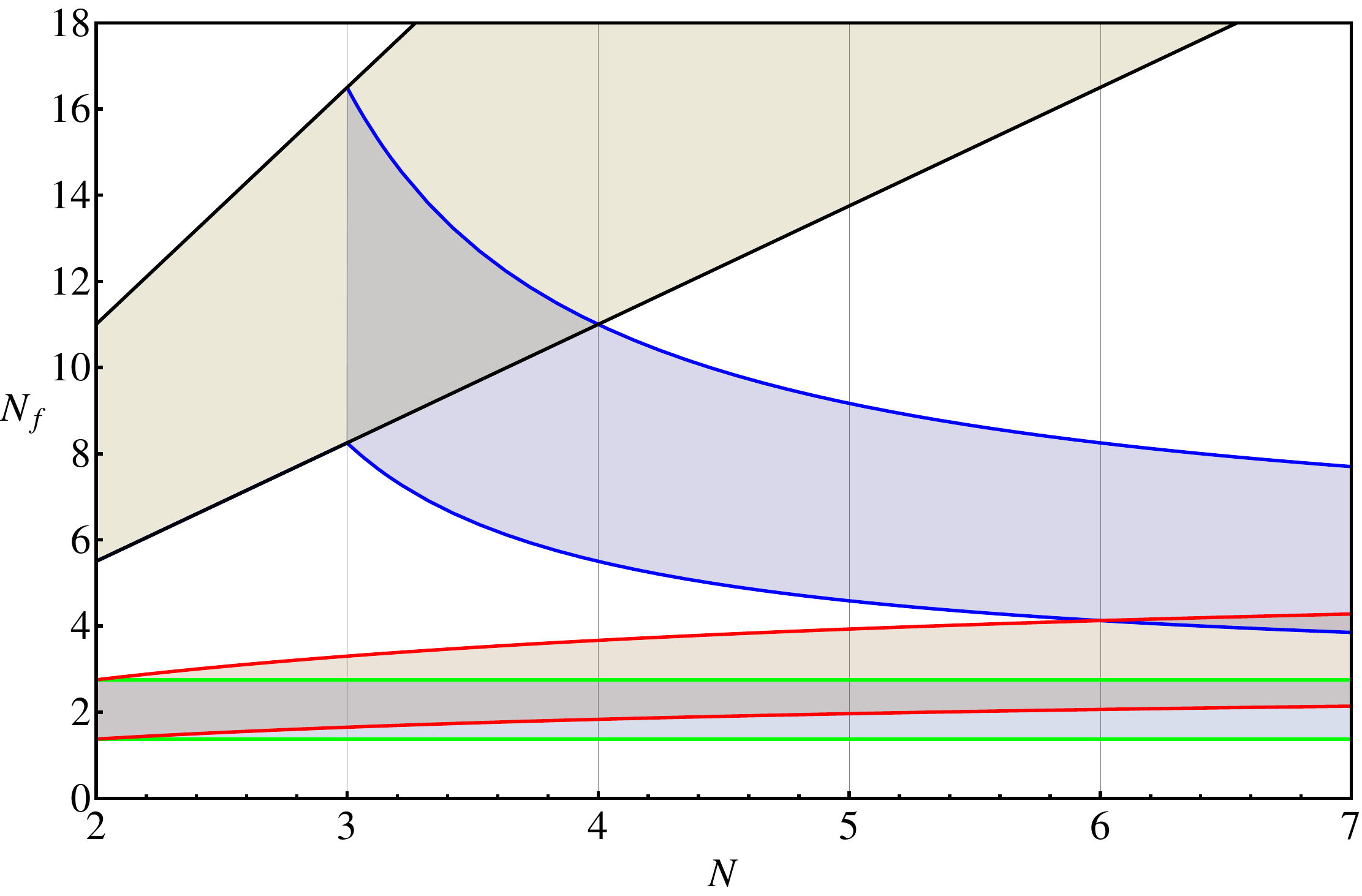}}
\caption{Phase diagram for theories containing fermions in the i) fundamental representation (black), ii) two-indexed antisymmetric representation (blue), iii) two-indexed symmetric representation (red), iv) adjoint representation (green). The shaded area is the conformal window} \label{PD}
\end{figure}

First the loss of asymptotic freedom is determined by the change of sign in the first coefficient of the beta function
\begin{eqnarray}
\sum_{i=1}^{k} \frac{4}{11} T(r_i) N_f(r_i) = C_2(G) \ .
\end{eqnarray}
Second we note that at the zero of the beta function we have
\begin{eqnarray}
\sum_{i=1}^{k} \frac{2}{11} T(r_i) N_f(r_i) \left(  2 + \gamma_i\right) = C_2(G) \ .
\end{eqnarray}
Having reached the zero of the beta function the theory is conformal in the infrared and hence the dimension of the chiral condensate must be larger than one in order not to contain negative norm states \cite{Mack:1975je}. Since the dimension of the chiral condensate is $3-\gamma_i$ we see that $\gamma_i =2$ for all representations $r_i$ yields the maximum possible bound of the conformal window
\begin{eqnarray}\label{bound}
\sum_{i=1}^{k} \frac{8}{11} T(r_i) N_f(r_i) = C_2(G) \ .
\end{eqnarray}
Hence in the case of a single representation the bound is
\begin{eqnarray}
\frac{11}{8} \frac{C_2(G)}{T(r)} < N_f(r) < \frac{11}{4} \frac{C_2(G)}{T(r)} \ .
\end{eqnarray}
In Fig. \ref{PD} we plot the conformal window for various representations.
One should note the remarkable feature that only a low number of flavors for the adjoint and two-indexed symmetric representation is needed in order to be near the conformal window. This has two important implications
\begin{itemize}
\item Such (near) conformal theories are easily accessible on the lattice.
\item They are perfect candidates for walking technicolor theories able to break the electroweak symmetry.
\end{itemize}
We stress that the above prediction of the conformal window is in agreement with all of the recent lattice calculations \cite{Catterall:2007yx}.

\section{Minimal Walking Technicolor}

From Fig. \ref{PD} it is clear that the simplest model able to possess walking dynamics is an $SU(2)$ gauge theory with two Dirac flavors in the adjoint representation. To couple it to the SM we arrange the left handed fields into three doublets of the $SU(2)_L$ weak interactions while the right handed fields are singlets under the SM gauge group. They are denoted by $Q^a_L =(U^a,D^a)_L\ ,U^a_R\ ,D^a_R\ ,a=1,2,3$.

The model so far suffers from the Witten topological anomaly \cite{Witten:1982fp}. This is easily accommodated for by adding a new fermionic doublet charged under the electroweak symmetry and neutral under the technicolor interactions $L_L = (N,E)_L\ ,N_R,E_R$. The gauge anomalies cancel for the following generic choice of hypercharge
\begin{eqnarray}
Y(Q_L) &=& \frac{y}{2} \ , \qquad Y\left( U_R,D_R \right) = \left( \frac{y+1}{2}, \frac{y-1}{2}  \right) \\
Y(L_L) &=& \frac{-3y}{2} \ ,\ \  Y\left( N_R,E_R \right) = \left( \frac{-3y+1}{2}, \frac{-3y-1}{2}  \right) \ ,
\end{eqnarray}
where $y$ can be any real number. The above model is called the Minimal Walking Technicolor (MWT) model \cite{Sannino:2004qp}. Once the condensate $\langle \overline{U}_R U_L + \overline{D}_R D_L \rangle$ forms the electroweak symmetry breaks providing masses for the associated gauge bosons.

Since the techniquarks belong to the adjoint representation of the gauge group the global symmetry is enhanced to $SU(4)$. Assuming the standard breaking to the maximal diagonal subgroup $SU(4)$ breaks to $SO(4)$. This leaves nine Goldstone bosons. Three of these become the longitudinal degrees of freedom of the massive weak gauge bosons while the remaining six Goldstone bosons acquire mass through yet unspecified extended technicolor interactions.

\section{Ultra Minimal Walking Technicolor}

Another possibility of constructing realistic walking technicolor models is to consider fermions transforming according to two distinct representations of the gauge group. First we are interested in having the smallest possible naive $S$ parameter. This is achieved by choosing two technicolors and two Dirac fermions in the fundamental representation. We charge these fermions under the electroweak symmetry in the standard way as done above for MWT. Second we are interested in obtaining walking dynamics. One solution is to add the remaining fundamental fermions uncharged under the electroweak symmetry needed to be near the conformal window. Such models have been termed partially gauged technicolor \cite{Dietrich:2006cm}.

However a more economic alternative is to let the remaining fermions belong to the adjoint representation of the gauge group. Then according to the prediction of the conformal window, Eq. (\ref{bound}), the critical number of adjoint Dirac flavors needed to enter the conformal window is $\sim 1$ depending on the critical value of the anomalous dimension. Hence our candidate theory consists of an $SU(2)$ gauge group with two Dirac flavors in the fundamental representation and charged under the electroweak symmetry together with one Dirac flavor in the adjoint representation uncharged under the electroweak symmetry. This model has been termed the Ultra Minimal Walking Technicolor (UMT) model \cite{Ryttov:2008xe}.

Due to the fact that the fermions belong to pseudoreal and real representations the global symmetry is enhanced to $SU(4)\times SU(2) \times U(1)$. All the fermions are charged under the abelian $U(1)$ symmetry which is anomaly free. Again assuming the standard breaking to the maximal diagonal subgroup the global symmetry breaks to $Sp(4) \times SO(2) \times Z_2$ leaving $5+2+1=8$ Goldstone bosons. Except for the triplet of Goldstone bosons which will be eaten by the massive gauge bosons the rest of the states are electroweak singlets. Specifically one of these states is a natural cold Dark Matter candidate whose mass can be very low \cite{Ryttov:2008xe}.


\begin{thebibliography}{199}

\bibitem{Weinberg:1979bn}
  S.~Weinberg,
  Phys.\ Rev.\  D {\bf 19}, 1277 (1979).
  L.~Susskind,
  Phys.\ Rev.\  D {\bf 20}, 2619 (1979).


\bibitem{Sannino:2008ha}
  F.~Sannino,
  arXiv:0804.0182 [hep-ph] and references therein.




\bibitem{Peskin:1990zt}
  M.~E.~Peskin and T.~Takeuchi,
  Phys.\ Rev.\ Lett.\  {\bf 65}, 964 (1990).
  M.~E.~Peskin and T.~Takeuchi,
  Phys.\ Rev.\  D {\bf 46}, 381 (1992).


\bibitem{Holdom:1981rm}
  B.~Holdom,
  Phys.\ Rev.\  D {\bf 24}, 1441 (1981).
  K.~Yamawaki, M.~Bando and K.~i.~Matumoto,
  Phys.\ Rev.\ Lett.\  {\bf 56}, 1335 (1986).
  T.~W.~Appelquist, D.~Karabali and L.~C.~R.~Wijewardhana,
  Phys.\ Rev.\ Lett.\  {\bf 57}, 957 (1986).
  T.~Appelquist and F.~Sannino,
  Phys.\ Rev.\  D {\bf 59}, 067702 (1999)
  [arXiv:hep-ph/9806409].




\bibitem{Sannino:2004qp}
  F.~Sannino and K.~Tuominen,
  Phys.\ Rev.\ D {\bf 71} (2005) 051901
  [arXiv:hep-ph/0405209].

\bibitem{Dietrich:2006cm}
  D.~D.~Dietrich and F.~Sannino,
  Phys.\ Rev.\  D {\bf 75}, 085018 (2007)
  [arXiv:hep-ph/0611341].

\bibitem{Ryttov:2007sr}
  T.~A.~Ryttov and F.~Sannino,
  Phys.\ Rev.\  D {\bf 76}, 105004 (2007)
  [arXiv:0707.3166 [hep-th]].

\bibitem{Ryttov:2007cx}
  T.~A.~Ryttov and F.~Sannino,
  Phys.\ Rev.\  D {\bf 78}, 065001 (2008)
  [arXiv:0711.3745 [hep-th]].

\bibitem{Ryttov:2008xe}
  T.~A.~Ryttov and F.~Sannino,
  Phys.\ Rev.\  D {\bf 78}, 115010 (2008)
  [arXiv:0809.0713 [hep-ph]].



\bibitem{Foadi:2007ue}
  R.~Foadi, M.~T.~Frandsen, T.~A.~Ryttov and F.~Sannino,
  Phys.\ Rev.\  D {\bf 76}, 055005 (2007)
  [arXiv:0706.1696 [hep-ph]].
  D.~D.~Dietrich, F.~Sannino and K.~Tuominen,
  Phys.\ Rev.\  D {\bf 72}, 055001 (2005)
  [arXiv:hep-ph/0505059]; {\it ibid} Phys.\ Rev.\ D {\bf 73} (2006) 037701
  [arXiv:hep-ph/0510217].
  A.~Belyaev, R.~Foadi, M.~T.~Frandsen, M.~Jarvinen, F.~Sannino and A.~Pukhov,
  arXiv:0809.0793 [hep-ph].
  O.~Antipin and K.~Tuominen,
  arXiv:0901.4243 [hep-ph].
  R.~Foadi, M.~Jarvinen and F.~Sannino,
  arXiv:0811.3719 [hep-ph].
  R.~Foadi and F.~Sannino,
  Phys.\ Rev.\  D {\bf 78}, 037701 (2008)
  [arXiv:0801.0663 [hep-ph]].
  F.~Sannino,
  arXiv:0811.0616 [hep-ph].
  D.~D.~Dietrich and M.~Jarvinen,
  arXiv:0901.3528 [hep-ph].
  D.~K.~Hong, S.~D.~H.~Hsu and F.~Sannino,
  Phys.\ Lett.\  B {\bf 597}, 89 (2004)
  [arXiv:hep-ph/0406200].
  N.~D.~Christensen and R.~Shrock,
  Phys.\ Lett.\  B {\bf 632}, 92 (2006)
  [arXiv:hep-ph/0509109].


\bibitem{Gudnason:2006mk}
  S.~B.~Gudnason, T.~A.~Ryttov and F.~Sannino,
  Phys.\ Rev.\  D {\bf 76}, 015005 (2007)
  [arXiv:hep-ph/0612230].
  N.~Chen and R.~Shrock,
  Phys.\ Rev.\  D {\bf 78}, 035002 (2008)
  [arXiv:0805.3687 [hep-ph]].



\bibitem{Cline:2008hr}
  J.~M.~Cline, M.~Jarvinen and F.~Sannino,
  Phys.\ Rev.\  D {\bf 78}, 075027 (2008)
  [arXiv:0808.1512 [hep-ph]].
  M.~Jarvinen, T.~A.~Ryttov and F.~Sannino,
  arXiv:0901.0496 [hep-ph].



\bibitem{Nardi:2008ix}
  E.~Nardi, F.~Sannino and A.~Strumia,
  arXiv:0811.4153 [hep-ph].
  R.~Foadi, M.~T.~Frandsen and F.~Sannino,
  arXiv:0812.3406 [hep-ph].
  S.~B.~Gudnason, C.~Kouvaris and F.~Sannino,
  Phys.\ Rev.\  D {\bf 74}, 095008 (2006)
  [arXiv:hep-ph/0608055].
  K.~Kainulainen, K.~Tuominen and J.~Virkajarvi,
  Phys.\ Rev.\  D {\bf 75}, 085003 (2007)
  [arXiv:hep-ph/0612247].
  C.~Kouvaris,
  Phys.\ Rev.\  D {\bf 76}, 015011 (2007)
  [arXiv:hep-ph/0703266].
  M.~Y.~Khlopov and C.~Kouvaris,
  Phys.\ Rev.\  D {\bf 77}, 065002 (2008)
  [arXiv:0710.2189 [astro-ph]].




\bibitem{Catterall:2007yx}
  S.~Catterall and F.~Sannino,
  Phys.\ Rev.\  D {\bf 76}, 034504 (2007)
  [arXiv:0705.1664 [hep-lat]].
  L.~Del Debbio, M.~T.~Frandsen, H.~Panagopoulos and F.~Sannino,
  JHEP {\bf 0806}, 007 (2008)
  [arXiv:0802.0891 [hep-lat]].
  S.~Catterall, J.~Giedt, F.~Sannino and J.~Schneible,
  JHEP {\bf 0811}, 009 (2008)
  [arXiv:0807.0792 [hep-lat]].
  Y.~Shamir, B.~Svetitsky and T.~DeGrand,
  Phys.\ Rev.\  D {\bf 78}, 031502 (2008)
  [arXiv:0803.1707 [hep-lat]].
  Y.~Shamir, B.~Svetitsky and T.~DeGrand,
  Phys.\ Rev.\  D {\bf 78}, 031502 (2008);  {\it ibid}
  arXiv:0812.1427 [hep-lat].
  L.~Del Debbio, A.~Patella and C.~Pica,
  arXiv:0805.2058 [hep-lat].
  A.~Hietanen, J.~Rantaharju, K.~Rummukainen and K.~Tuominen,
  PoS {\bf LATTICE2008}, 065 (2008); {\it ibid}
  arXiv:0812.1467 [hep-lat].
  T.~Appelquist, G.~T.~Fleming and E.~T.~Neil,
  Phys.\ Rev.\ Lett.\  {\bf 100}, 171607 (2008).
  A.~Deuzeman, M.~P.~Lombardo and E.~Pallante,
  arXiv:0804.2905 [hep-lat].
  Z.~Fodor, K.~Holland, J.~Kuti, D.~Nogradi and C.~Schroeder,
  arXiv:0809.4890 [hep-lat].
  X.~Y.~Jin and R.~D.~Mawhinney,
  PoS {\bf LATTICE2008}, 059 (2008)







\bibitem{Mack:1975je}
  G.~Mack,
  Commun.\ Math.\ Phys.\  {\bf 55} (1977) 1.

  M.~Flato and C.~Fronsdal,
  Lett.\ Math.\ Phys.\  {\bf 8}, 159 (1984).

  V.~K.~Dobrev and V.~B.~Petkova,
  Phys.\ Lett.\  B {\bf 162} (1985) 127.



\bibitem{Witten:1982fp}
  E.~Witten,
  Phys.\ Lett.\  B {\bf 117} (1982) 324.






\end{thebibliography}
\end{document}